\documentstyle[12pt]{article}
\begin{document}
\rightline {OSU-292}
\rightline {DOE-ER 40757-059}
\rightline{CPP-94-36-059}
\bigskip\bigskip\bigskip
\begin{center}
{\large \bf TOP QUARK SIGNATURE IN \\ EXTENDED COLOR THEORIES}\\
\bigskip\bigskip
D.A.Dicus$^{\dagger}$, B.Dutta$^{*}$ and S.Nandi$^{*}$\\
$\dagger $ Center for Particle Physics\\ The University of Texas at Austin\\
Austin, Texas 78712\\$*$Department of Physics\\Oklahoma State University\\
Stillwater, OK 74078
\end{center}
\clearpage
\begin{abstract}
We consider the implications of an extended color group, $SU\left( 3\right)
_I\times SU\left( 3\right) _{II}$, spontaneously broken to $SU\left(
3\right) _c$ at a TeV or lower scale, for the hadronic colliders. The
associated massive color octet gauge bosons (the colorons) can enhance the $t%
\overline{t}$ pair production at the Tevatron collider. At the LHC, the
colorons can be pair produced, each decaying to a $t\overline{t}$ pair. This
gives rise to anomalous multi W production: a clear signature of physics
beyond the Standard Model. We calculate the associated multijet and
multilepton final states at the Tevatron and the LHC energies, and compare
these with the expectations from the Standard Model.
\end{abstract}
\clearpage
\section{Introduction}

The CDF collaboration at the Fermilab Tevatron has reported \cite{[1]} the
observation of two charged dilepton and ten single lepton+$\geq 3$ jet
events which are in excess of those expected in the Standard model (SM)
excluding $t\overline{t}$ production. A detailed analysis of seven of these
events (which have at least one b-tag and a 4$^{th}$ jet) yields the central
value for top quark mass of 174 GeV and $t\overline{t}$ cross section of
13.9 pb at the Tevatron energy ($\sqrt{s}=1.8$ $TeV)$. This cross section is
about three times larger than expected in the Standard Model \cite{[2]}
although the error [1] is large. It is entirely possible that with larger
statistics, the values of the mass and the cross section will change to be
in agreement with the SM. However, it is also possible that we are seeing
the first glimpse of new physics beyond the SM at this TeV scale which is
being explored directly for the first time. Several ideas have been proposed
for new physics. One is to assume that the color group at high energy is
bigger, namely SU$\left( 3\right) _I\times $SU$\left( 3\right) _{II}$ \cite
{[3]}. The color I is coupled to the first two families of fermions while
the color II is coupled to the third family. This group breaks spontaneously
to the usual SU$\left( 3\right) _c$ at a TeV scale or below giving rise to
eight massive color octet gauge bosons, called colorons. Due to mixing,
these colorons couple to both the ordinary light quarks and to $t\overline{t}
$. These colorons are then produced from the ordinary light $q\overline{q}$
as resonances which then decay to $t\overline{t}$, thus enhancing the $t%
\overline{t}$ production. The second idea assumes the multiscale models of
walking technicolor\cite{[4]}. The color octet technipion, $\eta _T$ is
produced as a resonance in the gluon gluon channel and decay dominantly to $t%
\overline{t}$, thus increasing the $t\overline{t}$ production to the level
observed by CDF. In the third scenario, a singlet vector like, charge $%
+\frac 23$quark is assumed with a mass comparable to the top quark \cite{[5]}%
. This singlet quark mixes with the top. Their production and the subsequent
decay then effectively double the standard top signals [5]. Another idea
proposed is that the top quark may have anomalous chromomagnetic moment type
tree level coupling with the gluons \cite{[6]}. A small value of the
chromomagnetic moment, $\chi ,$can produce a cross section of the level
observed by the CDF Collaboration[6].

In this work, we discuss the hadronic collider implications of the first
idea above, an extended color model, $SU\left( 3\right) _I\times SU\left(
3\right) _{II}$, where the first two families of quarks couple to the $%
SU\left( 3\right) _I$ whereas the third family couples to $SU\left( 3\right)
_{II}$, as proposed in reference (3) .We calculate the multijet and / or
multilepton final state cross sections arising from production and the
subsequent decay of the coloron at the Fermilab Tevatron and Large Hadron
Collider(LHC) energies, and compare those with the expectations from the
Standrad Model. Hill and Parke have studied the coloron procuction at the
Fermilab Tevatron energy. At the Tevatron, the coloron is singly produced by
$q\overline{q}$ annihilation. There is no contribution from gluon-gluon
fusion, since there is no gluon-gluon-coloron coupling in this model. Hill
and Parke showed that for the extended color symmerty breaking scale at a
TeV or less, the resonant enhancement of the coloron production and their
subsequent decay to $t\overline{t}$ is enough to produce the large cross
section observed by the CDF collaboration. They also study the W and top
quark $p_T$ distributions and the $t\overline{t}$ mass distributions and
note that the larger $p_T$ in this model can be used to distinguish it from
the Standard Model. In this work we go further by looking at the decay
products of the W's and making some simple visiblity cuts to test the extent
to which the $p_T$ distributions as they would be observed, are really
different. However, the main part of our work is to study the implications
of the model at the LHC energy ($pp,\sqrt{s}=14TeV)$. Here, the colorons can
be pair produced via gluon-gluon fusion. Each coloron decays to a $t%
\overline{t}$ or $b\overline{b}$ pair. If we look at the tops, we get two
top quarks and two top antiquarks whose decays give rise to four W bosons in
the final state. The cross sections for these four W final states are much
larger than those in the Standard model. This anomalous W productions will
be very clean signal for physics beyond the Standard Model at high energy
hadronic colliders such as LHC. We also calculate the branching ratios for
the various mutijet and/or multilepton final state arising from the
subsequent decays of these final state W's.

We present our work as follows. In Sec.II, we give the formalism for the $%
SU\left( 3\right) _I\times SU\left( 3\right) _{II}$ extended color model and
write down the relevent interactions. In Sec.III, we discuss our results for
the Fermilab Tevatron. Sec.IV contains the main part of our work. In part A,
we calculate the differential and total cross sections for $pp\rightarrow $
coloron + coloron+ anything at various center of mass energies. In part B,
we discuss the topology of the events and calculate the branching ratios of
the multijet and/or multilepton final states by applying appropriate cuts.
Here we discuss our results and point out the inportant channels for clear
signals beyond the Standard Model. SectionV contains our conclusions.

\section{FORMALISM FOR THE $SU\left( 3\right) _I\times SU\left( 3\right)
_{II}$ COLOR MODEL}

The gauge part of the of the $SU\left( 3\right) _I\times SU\left( 3\right)
_{II}$ extended color model is

\begin{equation}
\label{one}-L_{gauge}=\frac 14F_{I\mu \upsilon a}F_{Ia}^{\mu \upsilon
}+\frac 14F_{II\mu \upsilon a}F_{IIa}^{\mu \upsilon }
\end{equation}
where

$$
F_{I\mu \upsilon a}=\partial _\mu A_{I\nu a}-\partial _\nu A_{I\mu
a}-h_1f_{abc}A_{I_{\mu b}}A_{I\nu c}
$$
and similarly for $F_{II\mu \nu a}$ with $h_1$replaced by $h_2$. $%
h_1$ and $h_2$ are the two color gauge coupling constants. The $SU\left(
3\right) _I\times SU\left( 3\right) _{II}$ symmetry is broken spontaneously
to the usual $SU\left( 3\right) _c$ at some scale M at or below a TeV. This
is achieved by using a Higgs field, $\Phi $ which fransform like (1,3,$%
\overline{3})$ under ($SU\left( 2\right) _L,SU\left( 3\right) _I,SU\left(
3\right) _{II})$ with VEV=diag.(M,M,M). At low energy, we are left with
eight massless gluons (A$_{\mu a}$) and eight massive colorons (B$_{\mu a})$
defined as

$$
A_I=A\cos \theta -B\sin \theta
$$

\begin{equation}
\label{two}A_{II}=A\sin \theta +B\cos \theta
\end{equation}
where $\theta $ is the mixing angle, and

\begin{equation}
\label{three}g_3=h_1\cos \theta =h_2\sin \theta \cdot
\end{equation}

The mass of the coloron is

\begin{equation}
\label{four}M_B=\left( \frac{2g_3}{\sin 2\theta }\right) M\cdot
\end{equation}

In terms of the gluon (A) and the coloron field (B), we can write the gauge
part of the interaction schematically as

\begin{equation}
\label{five}
\begin{array}{ll}
-L_{gauge}= & \frac 12g_3\left[ A^3+3AB^2+2\cot 2\theta
B^3\right] +\frac 14g_3^2[A^4+6A^2B^2+4\left( 2\cot 2\theta
\right) AB^3 \\  & +\left( \tan ^2\theta +\cot ^2\theta -1\right) B^4]\qquad
\cdot
\end{array}
\end{equation}

In Eq.(\ref{five}), A$^3$ and $A^4$ represent schematically the usual three
and four point gauge interactions, namaly,

$$
A^3\equiv f_{abc}(\partial _\mu A_{\nu a}-\partial _\nu A_{\mu a})A^{\mu
b}A^{\nu c}
$$
and

\begin{equation}
\label{six}A^4=f_{abc}f_{ade}A_{\mu b}A_{\nu c}A_d^\mu A_e^\nu \qquad\cdot
\end{equation}

We see from Eq.(\ref{five}) that a single coloron does not couple to two or
three gluons. Thus, a single coloron or a coloron in association with a
gluon can not be produced in hadronic colliders from gluon-gluon fusion.

The fermion representations under ($SU\left( 2\right) _L,SU\left( 3\right)
_I,SU\left( 3\right) _{II})$ are

$\left( u,d\right) _L,\left( c,s\right) _L\rightarrow \left( 2,3,1\right) ;$
$\left( u_R,d_R,c_R,s_R\right) \rightarrow $ $\left( 1,3,1\right) $

$\left( \nu _e,e\right) _L,\left( \nu _\mu ,\mu \right) ,\left( \nu _\tau
,\tau \right) _L\rightarrow \left( 2,1,1\right) ;$ $\left( e_R,\mu _R,\tau
_R,\nu _{iR}\right) \rightarrow \left( 1,1,1\right) $

\begin{equation}
\label{seven}\left( t,b\right) _L\rightarrow \left( 2,1,3\right) ;
\left( t_R,b_R\right) \rightarrow \left( 1,1,3\right) \cdot
\end{equation}

Note that the first two families of quarks couple to the color I while the
third family of quarks couples to color II. This assignment is anomaly free.
With the above assignment, the intersctions of all the quark with the gluons
are same as in the usual QCD. The interactions of the colorons are given by

\begin{equation}
\label{eight}-L_{coloron}=g_3\left[ z_1\sum_i\overline{q_i}\gamma ^\mu \frac{%
\lambda ^a}2q_i+z_2\left( \overline{t}\gamma ^\mu \frac{\lambda ^a}2t+%
\overline{b}\gamma ^\mu \frac{\lambda ^a}2b\right) \right] B_{\mu a}
\end{equation}
where the sum i is over u, d, s and c quarks.

\begin{equation}
\label{nine}z_1=-\tan \theta ,\qquad z_2=\cot \theta
\end{equation}
so that z$_1$z$_2=-1;$

\section{RESULTS FOR FERMILAB TEVATRON}

In this section, we discuss $t\overline{t}$ production and the resulting
multijet and/or multilepton final states at the $\overline{p}p$ collider at
the Fermilab Tevatron, $\sqrt{s}=1.8$ TeV. The dominant subprocess is the
annihilation of ordinary $q\overline{q}$ pair to produce $t\overline{t}$ via
coloron exchange in the s-channel, in addition to the usual standard model
processes. (The contribution of gg subprocess producing two colorons is
either kinematically not allowed or negligible). The total cross sections
depends on the coloron mass as well as the coloron width. We use Eq. (8) for
our calculations, and following Hill and Parke, present our results for the
coloron model, z$_1$z$_2=-1$ (Eq. 9), as well as for another model which is
like the coloron model except the value of z$_1$z$_2=+1$. [The value of z$_1$%
z$_2=+1$ can be obtained in a color singlet vector resonance model with an
extra U(1) \cite{[7]}]. For parton distributions, we use those produced by
the CTEQ collaboration\cite{[8]}.

Hill and Parke note that in their models the top quark and the W boson have
larger $p_T$ than in the Standard Model and that this could be used to
distinguish these models from the standard model with only a relatively
small number of top events. Here we go slightly further by looking at the
decay products of the Ws and making some simple visibility cuts.

In particular we keep the matrix element for top decay into bl$\overline{%
\upsilon }$ or $bq\overline{q}$ so as to include the coherent polarization
sum of the Ws. We then combine the quarks into jets by requiring that final
state quarks be in the same jet if their angular seperation is less than $%
\Delta R=0.5$ with the standard definition of $\Delta R$. We next require
that jets and the charged leptons from the Ws be visible by requiring that
their $p_T$ be larger than some $p_T^{\min }$ and that their rapidity $y$ be
less than some $y^{max}$. We also require that these leptons be seperated
from the jets, and from each other, if there is more than one, by $\Delta
R\geq 0.5.$

Using these criteria, we find the branching ratios for n jets and m charged
leptons where n=0,1,2,3,4,5,6 and m=0,1,2. We do this for the standard model
and for the following four models of Hill and Parke: (a) z$_1$z$_2=-1,$ M$%
_B=400GeV,$ $\Gamma _B=0.6$M$_B;$ $(b)$z$_1$z$_2==+1,$M$_B=600GeV,$ $\Gamma
_B=0.2$M$_B;$ $(c)$z$_1$z$_2=-1,$ M$_B=600GeV,$ $\Gamma _B=0.5$M$_B$ and (d)
z$_1$z$_2=+1,$ M$_B=400,$ $\Gamma _B=$M$_B.$

As can be seen from Fig.1, these four models each have a total cross section
near 14 pb. Table I gives the branching ratios if y$^{\max \qquad }$is $1.5$
and $p_T^{min}$ is 35 GeV. Clearly there is very little difference between
these models and the standard model (which is the top number in each set of
five) so far as BR's are concerned. Table II gives the same cases for $%
p_T^{min}=50$ GeV. Here the new models do show some $p_T$ behavior (except
for model (a)) but the branching ratios for the interesting topologies, four
jets and one lepton for example, are quite small. If the detector efficiency
is 10\% and we have 1000 pb$^{-1}$ of integrated luminosity then the
standard model gives 2.4 events of 4 jet, 1 lepton type, while the new
models give 4.2 to 23 events. Of course a large part of the extra events in
the new models is still just the larger cross section- 14 pb vs 5 pb for the
standard model.

We have also investigated other values of M$_B$ and $\Gamma _B/M_B$ and
found similar results. For $p_T^{min}=35GeV,$ the additional p$_T$ inherent
in these models is of only modest help in increasing the branching ratios of
the interesting topologies. For $p_T^{min}=50GeV$ the additional p$_T$ is a
big help but the branching ratios themselves are quite small.

We note that when we talk about charged leptons we mean electrons or muons.
We include the tau lepton by assuming it decays immmediately after
production into a muon or electron plus neutrinos (35.5\% of the time) or
into a quark pair plus a neutrino (64.5\% of the time). Thus the visible
final states of a W decay through a tau have the same particle content as
other W decays: an electron, a muon, or a pair of quarks. The possible
energies of the visible particles are, of course, different if the decay is
through a tau, and that has been included.

\section{COLORON SIGNAL AT LHC ENERGY}

In this section, we discuss the coloron pair productions in hadronic
collisions in the spontaneously broken $SU(3)_I\times SU(3)_{II}$ extended
color model. We consider only the case where each coloron decays to top
quarks, $t\overline{t}$. Decays of these t(or $\overline{t}$) to a W give
rise to four W bosons in the final state. The cross section for these four W
productions is much larger than that expected in the Standard model. This
anomalous W productions will be a very clean signal for physics beyond the
Standard Model. In section IVA, we calculate the differential and total
cross sections for the coloron pair production. In IVB, we discuss the
multijet and/or multilepton cross sections resulting from the decays of the
four tops produced from the coloron pairs.

\subsection{CROSS SECTIONS FOR COLORON PAIR \protect\newline PRODUCTIONS}

The dominant contribution to the coloron pair production at the LHC enengy
comes from the subprocess

\begin{equation}
\label{ten}g+g\rightarrow B+B
\end{equation}
The corresponding Feyman diagrams obtained from (5) are shown in Fig. 2. The
contribution of the other subprocess $q+\overline{q}\rightarrow B+B$ (Eq. 8)
is very small at the LHC energy because of the low q$\overline{q}$
luminosity.

The differential cross section for the subprocess (10) is obtained to be

\begin{equation}
\label{eleven}\frac{d\widehat{\sigma }}{dz}=\frac{9\pi \alpha _s^2}{512%
\widehat{s}}\beta F(\varepsilon ,z)
\end{equation}
where

\begin{equation}
\label{twelve}
\begin{array}{ll}
F\left( \epsilon ,z\right) = & [\frac 1{\left( 1+\beta z\right) ^2}(256+
\frac{48}{\epsilon ^2}) \\  & +\frac 1{(1+\beta z)}(-128-
\frac{96}\epsilon +\frac{24}{\epsilon ^2}) \\  & +(z\rightarrow -z)] \\
& +200+24\beta ^2z^2+\frac{48}\epsilon \qquad\cdot
\end{array}
\end{equation}
Here, $\widehat{s}$ is the total center of mass (CM) energy squared for the
subprocess, z is the cosine of the CM angle, M$_B$ is the mass of the
coloron, B, $\alpha _s$ is the QCD coupling constant squared over 4$\pi $,
and

\begin{equation}
\label{thirteen}\epsilon \equiv \frac{\widehat{s}}{4M_B^2},\qquad \beta
\equiv \left( 1-\frac 1\epsilon \right) ^{\frac 12}\qquad \cdot
\end{equation}
{}From Eq. (\ref{eleven}), we obtain the total subprocess cross section to be

\begin{equation}
\label{fourteen}
\begin{array}{ll}
\widehat{\sigma }= & \frac{9\pi \alpha _s^2}{512\widehat{s}}\beta \left[
\left( 1024\epsilon +416+\frac{272}\epsilon \right) -\left( 256+\frac{192}{
\epsilon} -\frac{48}{\epsilon ^2}\right) \frac 1\beta \ln \frac{1+\beta }{%
1-\beta }\right] \qquad\cdot
\end{array}
\end{equation}

The total cross section for the process

\begin{equation}
\label{fifteen}p+\overline{p}\rightarrow B+B+anything
\end{equation}
is obtained by folding in the gluon distributions with the above cross
section. We have used the distribution produced by the CTEQ collaboration
[8] evaluated at $Q^2=M_B^2$.

\subsection{MULTIJET-MULTILEPTON FINAL STATES \protect\newline FROM COLORON
PAIR DECAYS}

Using the production cross section above, and assuming the branching ratio
for each final states of W decay to be $\frac 19$, we find the branching
ratio for n jets and m charged leptons where n = 0,$\cdots ,12$ and m = 0,$%
\cdots ,4$. These results depend on the mass of the coloron, the visibility
cuts, and the mixing with light quarks (z$_1$ and z$_2$ in (8) above). To
isolate the combinations which can not be produced in lowest order of the SM
we consider only the $t\overline{t}$ $t\overline{t}$ final state.

As in the Fermilab results above, we combine the quarks into jets using the
condition that a quark belongs in an adjacent jet if the angular seperation
between them is less than $\Delta R$ which we take to be 0.5. Once the jets
are formed, we requires that their transverse momentum be larger than 30 GeV
and their rapidity be less 2. For the leptons (electron or muon) we require
a tranverse momentum of 20 GeV and a maximum rapidity of 2.5. The leptons
must be seperated from the jets, and from each other by $\triangle R$ which
is also taken to be 0.5. Thus, for example, if two leptons each satisfy the
transverse momentum and rapidity cuts but have\thinspace $\triangle R$ less
than 0.5 then they are counted as only one lepton.

If the final state of a W decay is a tau lepton, then we assume the tau has
decayed and use its decay products in forming jets and applying the
visibility cuts. In other words in case of a tau, we go one level further in
the decay chain to find the particles we treat as the final state.

We keep a coherent sum over the polarizations of the Ws from the top decays
but not for the Ws in the tau decays. We do not keep a coherent spin sum for
the tops.

Our results are given in Tables 3-5 for a coloron mass of 400, 600, and 800
GeV. We give results only for the branching ratios for events with more jets
or leptons than can be straight -forwardl\.y produced in the SM, or by other
final states of the colorons, $t\overline{t}$ $b\overline{b}$ for example.
To include mixing with the lighter quarks each branching ratio should be
miltiplied by

$$
\left[ \frac{z_2^2I}{4z_1^2+z_2^2+z_2^2I}\right] ^2
$$
where I is given by

$$
I=\left( 1+2\frac{m_t^2}{M_B^2}\right) \left( 1-4\frac{m_t^2}{M_B^2}\right)
^{\frac 12}\cdot
$$
These branching ratios are rather small; fortunately the production cross
sections are large so the actual number of events with these topologies can
be large.

\section{CONCLUSION}

At Fermilab energies we have calculated the branching ratios for the various
final states possible from a $t\overline{t}$ pair produced through
resonant coloron. This is quite model dependent; however, the motivation for
these model is that they can give a larger cross section than the SM. Thus
we choose sets of parameters which give a production cross section of about
three times the SM cross section. Our results are shown in Table 1 and 2. We
see that the branching ratios for the interesting final states are not very
different from those of the standard model unless we take a large value for
the minimum p$_T$. This seems to be in agreement with Ref. 3. We have chosen
to require a large transverse momentum for the jets and leptons because it
was hoped that these models would be distinguished by larger branching
ratios for large transverse momentum.

The only check we have on the accuracy of the numbers in any of the tables
is to rerun the Monte Carlo integral which generates the histograms with
different sets of random numbers. When we do this the branching ratios, even
those whose values are very small, remain very stable; nevertheless we feel
the very small numbers should not be trusted.

At LHC energies we have calculated the branching ratios for the various
final states of coloron-coloron$\rightarrow t\overline{t}$ $t\overline{t}$
production. These are given in Tables 3, 4 and 5. Here we have many final
states that are only possible in higher order in the SM; if n is the number
of jets and m is the number of electron or muons these final states are $m>2$
if $n\leq 2,$ $m>1$ if $3\leq n\leq 4,m>0$ if $5\leq n\leq 6$. Even when the
branching ratios for these states are rather small, this is compensated by a
large production cross section if the coloron mass is not too large.
Detection of these states would be a very strong signal for the coloron.

\newpage\

\section{Acknowledgements}

This work was supported in part by the U.S. Department of Energy Grants
DE-F603-93ER 40757 and DE-FGO2-94ER 40852.

\newpage\

\newpage\

{\bf FIGURE CAPTION}

\begin{quote}
{\bf Figure 1}. Feynman diagrams for the process gluon+gluon $\rightarrow $
coloron+coloron.

{\bf Figure 2.} Cross sections (in pb) for the $t\overline{t}$ pair
productions at the Tevatron ($\sqrt{s}=1.8TeV).$ $M_{B}$ and $\Gamma
_B$ are the mass and the width of the coloron. The solid curves are for $%
z_1z_2$ = -1 white the dotted curves are $z_1z_2$ = $+$1 as discussed in
Section III. The numbers indicated with the curves are the coloron masses in
GeV. The four models discussed in the text are indicated by (a), (b),(c) and
(d). The experimental value of the cross section, as measured by CDF
collaboration, is shown by the arrow.
\end{quote}

\newpage\

{\bf TABLE CAPTION}

\begin{quote}
{\bf Table 1.}

Branching ratios for the various multijet and multilepton final states with
each jet and charged lepton (e or $\mu )$ having $p_{_T}>35$ GeV and with
other cuts as discussed in the text. The decays of $\tau $ to e, $\mu $ or
quarks have been included. The results are for the Tevatron energy, $\sqrt{s%
}$ =1.8 TeV. SM stands for the Standard Model and (a), (b), (c) and
(d) are the four different models discussed in Section 3.

{\bf Table 2. }

Same as in table 1 except for $p_{_T}>50$ GeV.

{\bf Table 3}.

Branching ratios for the various multijet and multilepton (e or $\mu )$
final states at the LHC energy, $\sqrt{s}$ =14 TeV for the coloron
model with coloron mass $M_B$= 400 GeV. The cuts are $\left(
p_T^{jets}\right) _{\min }=30$ $GeV,$ $\left( p_T^{leptons}\right) _{\min
}=20$ $GeV$, y$_{jet}\leq $ 2.0, y$_{lepton}\leq $ 2.5, and $\Delta R$ = 0.5
everywhere as discussed section 4.2. The top mass has been taken to be 175
GeV. The decays of $\tau $ to e, $\mu $ or quarks have been included. Each
branching ratio in the table should be multiplied by $[\frac
{0.6695}{4z_1^2/z_2^2+1.6695}]^2.$

\newpage\

{\bf Table 4}.

Same as in table 3, except for $M_B$= 600 GeV. Each branching ratio in the
table should be multiplied by $[\frac{0.9504}{4z_1^2/z_2^2+1.9504}]^2$\
{}.

{\bf Table 5}.

Same as in table 3 except for $M_B$= 800 GeV. Each branching ratio in the
table should be multiplied by $[\frac{0.9853}{4z_1^2/z_2^2+1.9853}]^2$ $%
\cdot $
\end{quote}

\renewcommand{\arraystretch}{.7}
\begin{table}
\centering
\caption{}
\begin{tabular}{l|cccc}
\hline
jets $\backslash$ leptons   &       &    0     &   1           &     2      \\

                           &   (SM) & 1.81E-3  & 2.32E-3    &  9.09E-4   \\

                          &   (a)   &  2.17E-3  & 2.14E-3    &  7.59E-4    \\

0                         &   (b)   &  2.00E-3  & 2.17E-3    &  1.27 E-3    \\

                          &   (c)   &  1.80E-3  & 2.20E-3    &  1.01 E-3
\\

                          &   (d)   &  1.95E-3  & 2.15E-3    &   9.93 E-3
\\\hline

                          &         &  0.0251   &  0.0245    &   5.93E-3
\\

                          &         &  0.0249   &  0.0259    &   5.48E-3
\\

1                         &         &   0.0191  &  0.0240    &   7.89E-3
\\

                          &         &   0.0231  &  0.0256    &   6.45E-3
\\

                          &         &   0.0229  &  0.0253    &   6.45E-3    \\
 \hline

                          &         &   0.115   &   0.0825    &   9.00E-3  \\

                          &         &   0.121   &   0.0810    &   8.53E-3   \\

2                         &         &   0.0874  &   0.0786    &   0.0124     \\

                          &         &   0.107   &   0.0819    &   9.92E-3
\\

                          &         &   0.106   &   0.0823    &   0.0104
\\ \hline

                          &         &   0.236   &   0.0897    &   \\

                          &         &   0.240   &   0.0847    &   \\

3                         &         &   0.188   &   0.106     &  \\

                          &         &   0.225   &   0.0938    &   \\

                          &         &   0.220   &   0.0942    &   \\  \hline

                          &         &   0.239   &   0.0298    & \\

                          &         &   0.239   &   0.0270    &   \\

4                         &         &   0.230   &   0.0473    &   \\

                          &         &   0.238   &   0.0334    &   \\

                          &         &   0.239   &   0.0361    &   \\ \hline

                          &         &   0.117   &             & \\

                          &         &   0.110   &             & \\

5                         &         &   0.152   &  &$\sigma$:4.810 $\pm$ 0.009
pb SM            \\

                          &         &   0.126   &      &    :13.93 $\pm$  0.03
pb  (a)             \\

                          &         &   0.128   &       &   :13.49 $\pm$ 0.04
pb   (b)              \\\cline{2-3}

                          &          &           &       &  :13.52 $\pm$ 0.02
pb   (c)            \\

                          &         &   0.0209   &      &   :13.89 $\pm$ 0.03
pb   (d)          \\

                          &         &   0.0177   &                 &\\

6                         &         &   0.0388   &                 & \\

                          &         &   0.0241   &                   &  \\

                          &         &   0.0263   &                   &
\\\hline
\end{tabular}
\end{table}
\begin{table}
\centering
\caption{}
\begin{tabular}{l|cccc}
\hline
jets $\backslash$ leptons   &       &    0     &   1           &     2      \\

                           &   (SM) &  0.0155  & 6.81E-3    &  1.36E-4   \\

                          &   (a)   &  0.0188  & 7.30E-3    &  1.31E-4    \\

0                         &   (b)   &  0.0107  & 5.70E-3    &  1.84 E-3    \\

                          &   (c)   &  0.0148  & 6.66E-3    &  1.53 E-3      \\

                          &   (d)   &  0.0137  & 6.85E-3    &   1.57 E-3
\\\hline

                          &         &  0.105   &  0.0430    &   3.48E-3
\\

                          &         &  0.109   &  0.0455    &   3.04E-3
\\

1                         &         &   0.0602  &  0.0395    &   6.04E-3
\\

                          &         &   0.0927  &  0.0428    &   3.98E-3
\\

                          &         &   0.0876  &  0.0437    &   4.32E-3    \\
 \hline

                          &         &   0.270   &   0.0670    &   2.50E-3  \\

                          &         &   0.295   &   0.0631    &   1.87E-3   \\

2                         &         &   0.177  &   0.0795    &     5.73E-3
\\

                          &         &   0.257   &   0.0696    &   2.85E-3
\\

                          &         &   0.250   &   0.0720    &   3.12E-3
 \\ \hline

                          &         &   0.280   &   0.0370    &   \\

                          &         &   0.290   &   0.0304    &   \\

3                         &         &   0.258   &   0.0684     &  \\

                          &         &   0.283   &   0.0409    &   \\

                          &         &   0.278   &   0.0447    &   \\  \hline

                          &         &   0.130   &   4.58E-3    & \\

                          &         &   0.121   &   2.48E-3  &   \\

4                         &         &   0.197   &   0.0161   &   \\

                          &         &   0.141   &   5.97E-3    &   \\

                          &         &   0.150   &   6.70E-3  &   \\ \hline

                          &         &   0.0262   &             & \\

                          &         &   0.0176   &             & \\

5                         &         &   0.0722   &             &  \\

                          &         &   0.0320   &             &   \\

                          &         &   0.0355   &             &   \\ \hline

                          &         &   2.29E-3  &         &  \\

                          &         &   7.31E-4   &                &\\

6                         &         &   9.62E-3   &                  & \\

                          &         &   3.14E-3   &                  &  \\

                          &         &   3.49E-3   &                   &
\\\hline
\end{tabular}
\end{table}
\begin{table}
\centering
\caption{}
\begin{tabular}{l|ccccc}
\hline
jets $\backslash$ leptons & 0       & 1        & 2       & 3       & 4  \\

0                         &         &          &         & 1.18E-5 & 1.80E-6 \\

1                         &         &          &         & 1.76E-4 & 3.55E-5
\\

2                         &         &          &         & 1.15E-3 & 1.67E-4 \\

3                         &         &          &         & 3.30E-3 & 3.50E-4 \\

4                         &         &          &         & 5.05E-3 & 2.43E-4 \\

5                         &         &          & 0.0326  & 3.86E-3 &  \\

6                         &         &          & 0.0258  & 1.04E-3 &  \\

7                         &         & 0.0770   & 0.0103  &        &  \\

8                         &         & 0.0391    & 1.69E-3    & \\

9                         &  0.0589 & 0.0106    &    &\\

10                        & 0.0192  & 1.11E-3  &           &$\sigma$ = 758
$\pm$ 2pb          &  \\

11                        & 4.01E-3  &  &           &         &  \\

12                        & 3.51E-4  &  &           &         &  \\
\end{tabular}
\end{table}
\begin{table}
\centering
\caption{}
\begin{tabular}{l|ccccc}
\hline
jets $\backslash$ leptons & 0       & 1        & 2       & 3       & 4  \\

0                         &         &          &         & 4.75E-6 & 1.37E-6 \\

1                         &         &          &         & 9.65E-5 & 2.45E-5
\\

2                         &         &          &         & 7.94E-4 & 1.66E-4 \\

3                         &         &          &         & 3.13E-3 & 4.29E-4 \\

4                         &         &          &         & 6.03E-3 & 3.65E-4 \\

5                         &         &          & 0.0347  & 5.40E-3 &  \\

6                         &         &          & 0.0335  & 1.86E-3 &  \\

7                         &         & 0.0942   & 0.0173  &         &  \\

8                         &         & 0.0607   & 3.73E-3 &
    &  \\

9                         & 0.0792  & 0.0207   &     &            &\\

10                        & 0.0350  &  3.22E-3 &   & $\sigma$=67.32 $\pm$
0.16pb        &  \\

11                        & 9.75E-3  &   &   &         &  \\

12                          & 1.09E-3  &   &   &         &  \\
\end{tabular}
\end{table}
\begin{table}
\centering
\caption{}
\begin{tabular}{l|ccccc}
\hline
jets $\backslash$ leptons & 0       & 1        & 2       & 3       & 4  \\

0                         &         &          &         & 1.71E-6 & 7.62E-7 \\

1                         &         &          &         & 6.02E-5 & 1.70E-5
\\

2                         &         &          &         & 5.92E-4 & 1.29E-4 \\

3                         &         &          &         & 2.76E-3 & 3.61E-4 \\

4                         &         &          &         & 5.51E-3 & 3.23E-4 \\

5                         &         &          & 0.0334  & 4.76E-3 &  \\

6                         &         &          & 0.0302  & 1.55E-3 &  \\

7                         &         & 0.0879   & 0.0142  &         &  \\

8                         &         & 0.0607   & 3.73E-3 &
    &  \\

9                         & 0.0687  & 0.016   &     &            &\\

10                        & 0.0287  &  2.29E-3 &   & $\sigma$=9.448 $\pm$
0.024pb        &  \\

11                        & 6.71E-3  &   &   &         &  \\

12                          & 8.79E-4  &   &   &         &  \\
\end{tabular}
\end{table}


\begin{thebibliography}{9}
\bibitem{[1]}  F. Abe et. al, Phys. Rev. Lett. \underline{73}, 225(1994);
ibid, Phys. Rev. \underline{D 50}, 2966 (1994).

\bibitem{[2]}  E. Laenen, J. Smith, and W.L.Van Neerven, Phys. Lett.
\underline{B321}, 254(1994).

\bibitem{[3]}  C.T. Hill, S.J. Parke, Phys. Rev. \underline{D 49,} 4454
(1994); C.T. Hill, Phys. Lett. \underline{B266}, 419 (1991).

\bibitem{[4]}  E. Eichten and K. Lane.
FERMILAB-PUB-94/007-T/BUHEP-94-1(1994); T.Appelquist and G.Triantaphyllou,
Phys. Rev. Lett. \underline{69}, 2750 (1992); E. Eichten, I. Hinchcliffe, K.
Lane and C. Quigg, Phys. Rev. \underline{D 34}, 1547 (1986).

\bibitem{[5]}  V. Barger and R.J.N. Phillips. University of Wisconsin
Preprint, MAD/PH/830 (1994), hep-ph 9405224; W. S. Hou and H. Huang,
National Taiwan University Preprint, NTUTH-94-18(1994) (hep-ph 9409227); W.
S. Hou, Phys. Rev. Lett. \underline{72}, 3945 (1994); B. Mukhopadhyaya and
S.Nandi, Phys. Rev. Lett. \underline{66}, 285 (1991); ibid; Phys. Rev.D.%
\underline{46}, 5098 (1992); T.P.Cheng and L.-F. Li, Phys. Rev \underline{D45%
}, 1708 (1992) W.S. Hou, Phys. Rev. Lett. \underline{69}, 3587(1992); B.
Mukhopadhyaya and S. Nandi, Phys. Rev. Lett. \underline{24}, 3588 (1992).

\bibitem{[6]}  D. Atwood, A. Kagan and T.G. Rizzo, SLAC-PUB-6580, July,
1994; G. L. Kane, G.A. Ladinsky and C.P. Yuan, Phys. Rev. D \underline{45},
124 (1992).

\bibitem{[7]}  M. Lindner and D. Ross, Nucl. Phys. B \underline{370}, 30
(1992).

\bibitem{[8]}  R. Brock et. al. (CTEQ Collaboration), ''Handbook of
Perturbative QCD, Version 1.0'', Fermilab-Pub-93-094 (1993).
\end{thebibliography}
\end{document}